\documentclass[preprint, twocolumn,5p,authoryear]{elsarticle}

\usepackage{amssymb}

\usepackage{graphicx}
\usepackage{subcaption}

\usepackage{amsmath}

\journal{ESWA}

\begin{document}

\begin{frontmatter}

\title{Dual Attention Driven Lumbar Magnetic Resonance Image Feature Enhancement and Automatic Diagnosis of Herniation}

\author[1]{Lingrui Zhang \fnref{label2}}
\author[2]{Liang Guo \fnref{label2}}
\author[2]{Xiao An}
\author[2]{Feng Lin}
\author[4]{Binlong Zheng \corref{cor1}} 
\author[1,3]{Jiankun Wang \corref{cor1}} 
\author[2]{Zhirui Li \corref{cor1}} 

\affiliation[1]{organization={Shenzhen Key Laboratory of Robotics Perception and Intelligence, and the Department of Electronic and Electrical Engineering, Southern University of Science and Technology},
	city={Shenzhen},
	postcode={518055}, 
	country={China}}

\affiliation[2]{organization={ Department of Orthopedics, Hainan Hospital of Chinese People's Liberation Army General Hospital},
	city={Sanya},
	postcode={572000}, 
	country={China}}

\affiliation[3]{organization={Jiaxing Research Institute, Southern University of Science and Technology},
	city={Jiaxing},
	postcode={314000}, 
	country={China}}

\affiliation[4]{organization={Shandong Juzhong Digital Medicine Technology Development Co.},
	city={Linyi},
	postcode={276100}, 
	country={China}}

\cortext[cor1]{Corresponding author.}
\fntext[label2]{These authors contributed equally.}

\begin{abstract}
Lumbar disc herniation (LDH) is a common musculoskeletal disease that requires magnetic resonance imaging (MRI) for effective clinical management. However, the interpretation of MRI images heavily relies on the expertise of radiologists, leading to delayed diagnosis and high costs for training physicians. Therefore, this paper proposes an innovative automated LDH classification framework. To address these key issues, the framework utilizes T1-weighted and T2-weighted MRI images from 205 people. The framework extracts clinically actionable LDH features and generates standardized diagnostic outputs by leveraging data augmentation and channel and spatial attention mechanisms. These outputs can help physicians make confident and time-effective care decisions when needed. The proposed framework achieves an area under the receiver operating characteristic curve (AUC-ROC) of 0.969 and an accuracy of 0.9486 for LDH detection. The experimental results demonstrate the performance of the proposed framework. Our framework only requires a small number of datasets for training to demonstrate high diagnostic accuracy. This is expected to be a solution to enhance the LDH detection capabilities of primary hospitals.
\end{abstract}

\begin{keyword}
Lumbar disc herniation, Magnetic resonance images, Deep learning
\end{keyword}

\end{frontmatter}

\section{Introduction}
The lumbar spine comprises five vertebrae (L1-L5) and is the human body's main load-bearing structure and the central trunk movement hub. The intervertebral discs absorb impact and protect the vertebral bodies. The body promotes flexion, extension, lateral bending, and rotation through the coordinated movement of the joint surfaces and intervertebral discs. The network of ligaments and muscles around the lumbar spine maintains stability and posture. The lumbar spine also contains the spinal canal, which protects the nerve roots in the conus medullaris and neural foramina. \citep{basak2025machine}. 

Clinically, the lumbar spine is susceptible to diseases such as LDH, spinal stenosis, and spondylolisthesis. These diseases are often related to repetitive strain or poor posture. Lumbar disc herniation is a common musculoskeletal disease. It occurs when the soft inner core of the intervertebral disc protrudes through a crack in the tough outer layer and compresses the adjacent nerve roots or spinal cord \citep{valarmathi2023automatic}. LDH condition most frequently involves lower lumbar levels. It often presents with symptoms such as radicular pain (shooting pain along the sciatic nerve), sensory deficits, and motor weakness in the lower extremities \citep{nisar2024lumbar}. Thin discs characterize the L4-L5 and L5-S1 segments. However, these segments are subject to extreme biomechanical demands, including concentration of up to 80\% of axial spinal loads and frequent dynamic movements. This results in a high prevalence of LDH in these two segments of 26.08\% and 24.09\%, respectively \citep{zhang2016identification}, \citep{ropper2015sciatica}.

In today's society, the age of LDH patients has gradually decreased. In 2006, the average age of LDH patients was around 45 years old \citep{hing2006national}. By 2016, the average age of the LDH incidence group had dropped to around 37 years old\citep{zhang2016identification}. A delayed diagnosis, whether due to misdiagnosis (Average misdiagnosis rate: 12\%-15\%) or other factors, can lead to persistent compression of the spinal cord or nerve roots. This prolonged compression may result in lumbar spondylolisthesis, intervertebral space infections, and other serious complications \citep{bevculic2016impact}, \citep{small2005orthopedic}.

\begin{figure}[htbp]
  \centering
  \begin{subfigure}[htbp]{0.45\textwidth}
    \includegraphics[width=\textwidth]{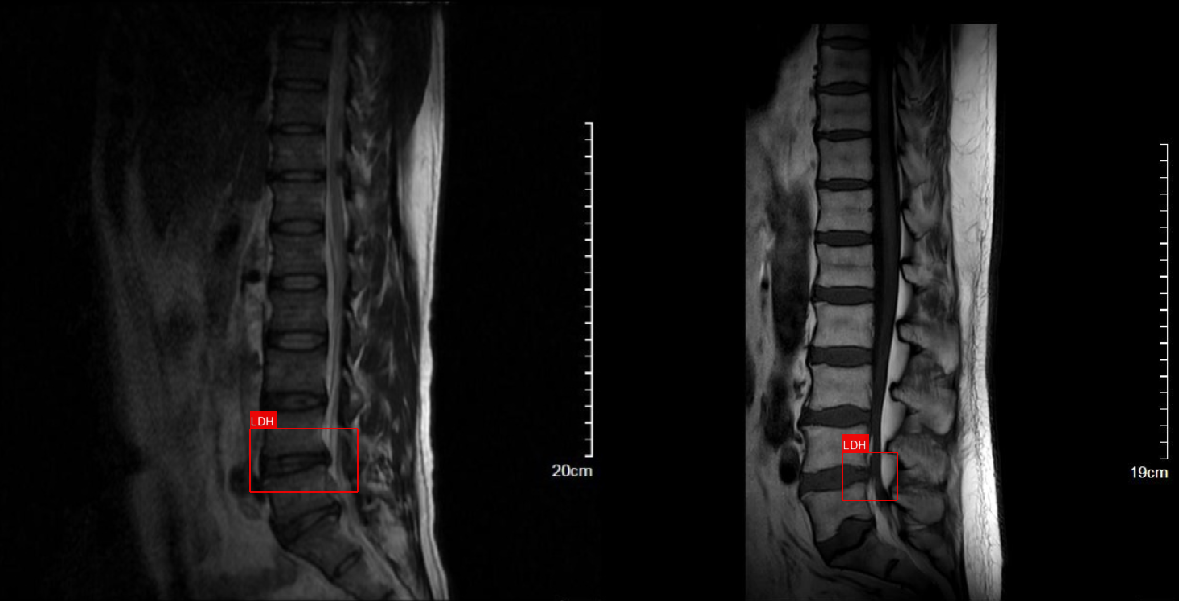}
    \caption{Sagittal Plane}
    \label{sub_Sagittal_Plane}
  \end{subfigure}
  \hfill
  \begin{subfigure}[htbp]{0.45\textwidth}
    \includegraphics[width=\textwidth]{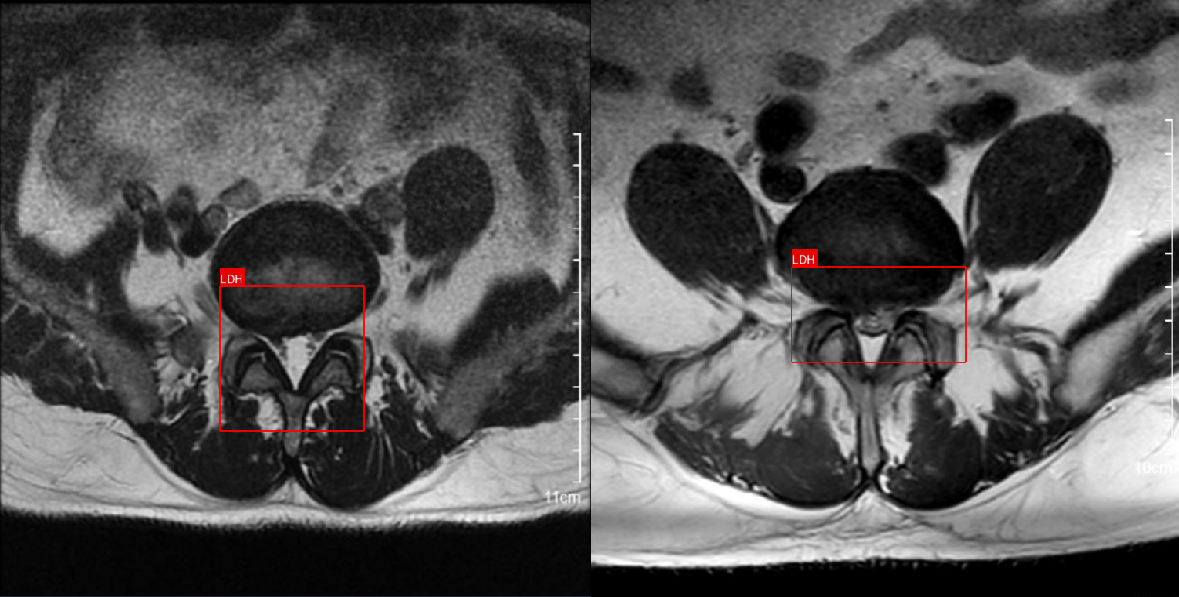}
    \caption{Axial Plane}
    \label{sub_Axial_Plane}
  \end{subfigure}
  \caption{MRI of LDH. (a), The Sagittal Plane of MRI. (b), The axial os plane of MRI. The red box indicates the location of the LDH lesion.}
  \label{fig_MRI}
\end{figure}

When determining LDH, a comprehensive assessment of the relationship between nerve root or cauda equina compression and disc degeneration is required. This requires combining information from the sagittal plane (Fig.\ref{sub_Sagittal_Plane}) and the axial plane (Fig.\ref{sub_Axial_Plane}). The Sagittal Plane is primarily used to determine the longitudinal severity of LDH. The presence of Bulging or Protrusion can be treated conservatively. The presence of Extrusion or Sequestration requires surgery. The Axial Plane is used to locate the transverse position of the herniation. This allows visualization of whether the herniation is Central or Para-Central to determine the degree of nerve compression and treatment options \citep{pfirrmann2001magnetic}. 

However, many methods for classifying LDH are trained using only the Sagittal Plane. Using a model that relies only on the Sagittal Plane may omit critical information, leading to inadequate clinical decision support and misdiagnosis or missed diagnosis \citep{sari2024enhanced}. Some existing methods use only YOLO and the convolutional neural network (CNN) \citep{prisilla2023approach}, \citep{tsai2021lumbar}, \citep{duan2024enhanced}. These methodologies exhibit limitations in accurately capturing subtle variations in lumbar spine structure during real-time detection and do not incorporate specialized optimizations tailored to the distinct anatomical features of the lumbar spine.

\begin{figure*}[htbp]
  \centering
  \begin{subfigure}[htbp]{0.3\textwidth}
    \includegraphics[width=\textwidth]{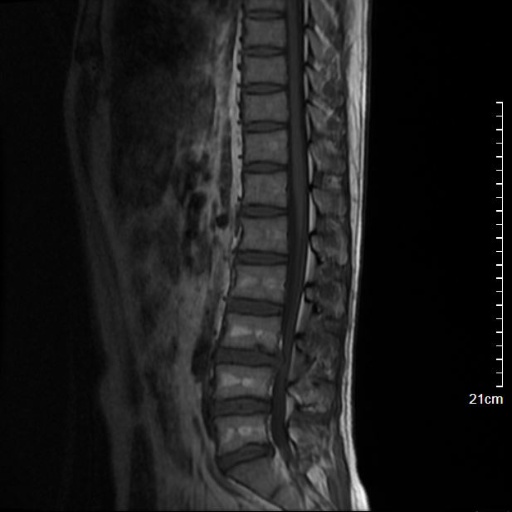}
    \caption{T1-weighted image of the Sagittal Plane}
    \label{sub_T1}
  \end{subfigure}
  \hfill
  \begin{subfigure}[htbp]{0.3\textwidth}
    \includegraphics[width=\textwidth]{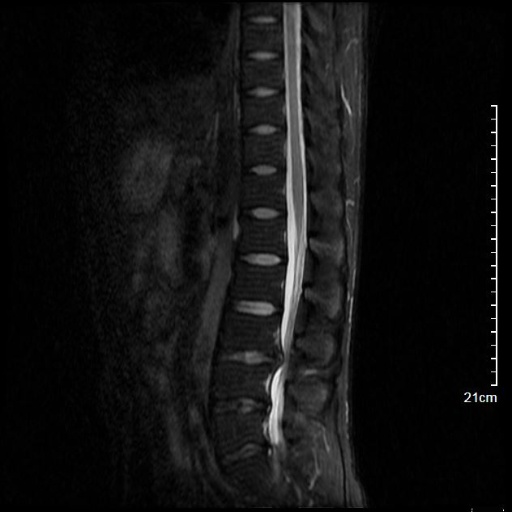}
    \caption{T2-weighted image of the sagittal plane}
    \label{sub_T2}
  \end{subfigure}
  \hfill
  \begin{subfigure}[htbp]{0.3\textwidth}
    \includegraphics[width=\textwidth]{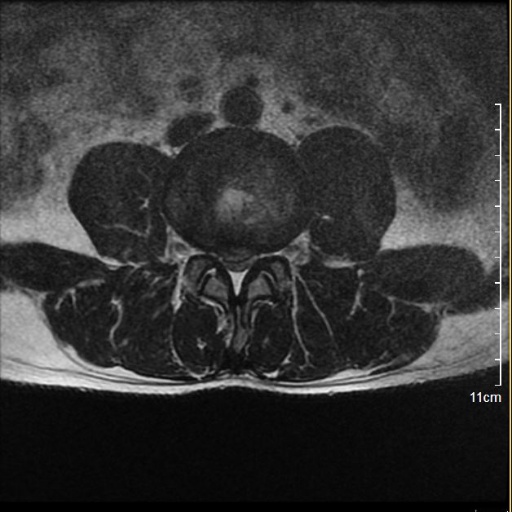}
    \caption{T2-weighted images of the Axial Plane}
    \label{sub_T2_A}
  \end{subfigure}
  \caption{Different types of MRI. (a), T1-weighted images of the Sagittal Plane, reflecting differences in longitudinal relaxation times of tissues, can clearly show anatomical structures. (b), T2-weighted images of the Sagittal Plane, reflecting differences in tissue transverse relaxation times, can be used to visualize soft tissue lesions. (c), T2-weighted images of the Axial Plane.}
  \label{fig_MRI_type}
\end{figure*}

This study presents a novel methodology integrating a self-attention-based neural network architecture to develop an automated classification system for LDH utilizing three distinct MRI (Fig.\ref{fig_MRI_type}) modalities—a framework demonstrating enhanced clinical relevance.  Our main contributions are as follows:
\begin{itemize}
    \item By leveraging the strengths of T1-weighted Sagittal Plane Images for clear anatomical visualization and T2-weighted Sagittal and Axial Plane Images' sensitivity to disc degeneration, herniation, and nerve root compression. We aim to equip healthcare providers with enhanced clinical decision-making capabilities when evaluating LDH.
    \item Using data augmentation techniques suitable for LDH, we minimize the requisite number of training samples, thereby mitigating annotation burdens and reducing overall training expenses.
    \item The integration of attention mechanisms allows the model to effectively capture subtle changes within the lumbar spine structure, resulting in a 97\% accuracy rate for our model.
\end{itemize}

\section{Related works}

Machine learning methodologies have revolutionized medical image processing, which is pivotal in enhancing diagnostic accuracy and clinical efficiency. By leveraging complex algorithms and deep learning architectures, these methods enable precise identification, segmentation, and classification of anatomical structures and pathologies within medical images \citep{latif2019medical}. This facilitates early detection of abnormalities and supports personalized treatment planning, ultimately improving patient outcomes and prognostic predictions. Integrating machine learning into medical imaging workflows represents a transformative advancement in healthcare delivery.

Over the past decades, efforts have been devoted to developing robust methods for the automated detection of LDH. Alomari et al. \citep{alomari2010computer} pioneered a probabilistic Gaussian framework rooted in the Gibbs distribution theory to classify patients even on limited datasets. Building on advances in image processing, Ghosh et al. \citep{ghosh2014supervised} pioneered an innovative hybrid approach combining the Hough transform algorithm \citep{deans1981hough} with adaptive windowing techniques to improve the detection accuracy of herniated disc features in medical imaging. He et al. \citep{he2016automated} introduced a pixel-centered analysis paradigm utilizing simultaneous hyperpixel segmentation to identify intervertebral foraminal stenosis and classify patients accurately. However, the accuracy of these methods does not meet modern standards for clinical use.

\begin{figure*}[htbp]
  \centering
  \includegraphics[width=0.9\textwidth]{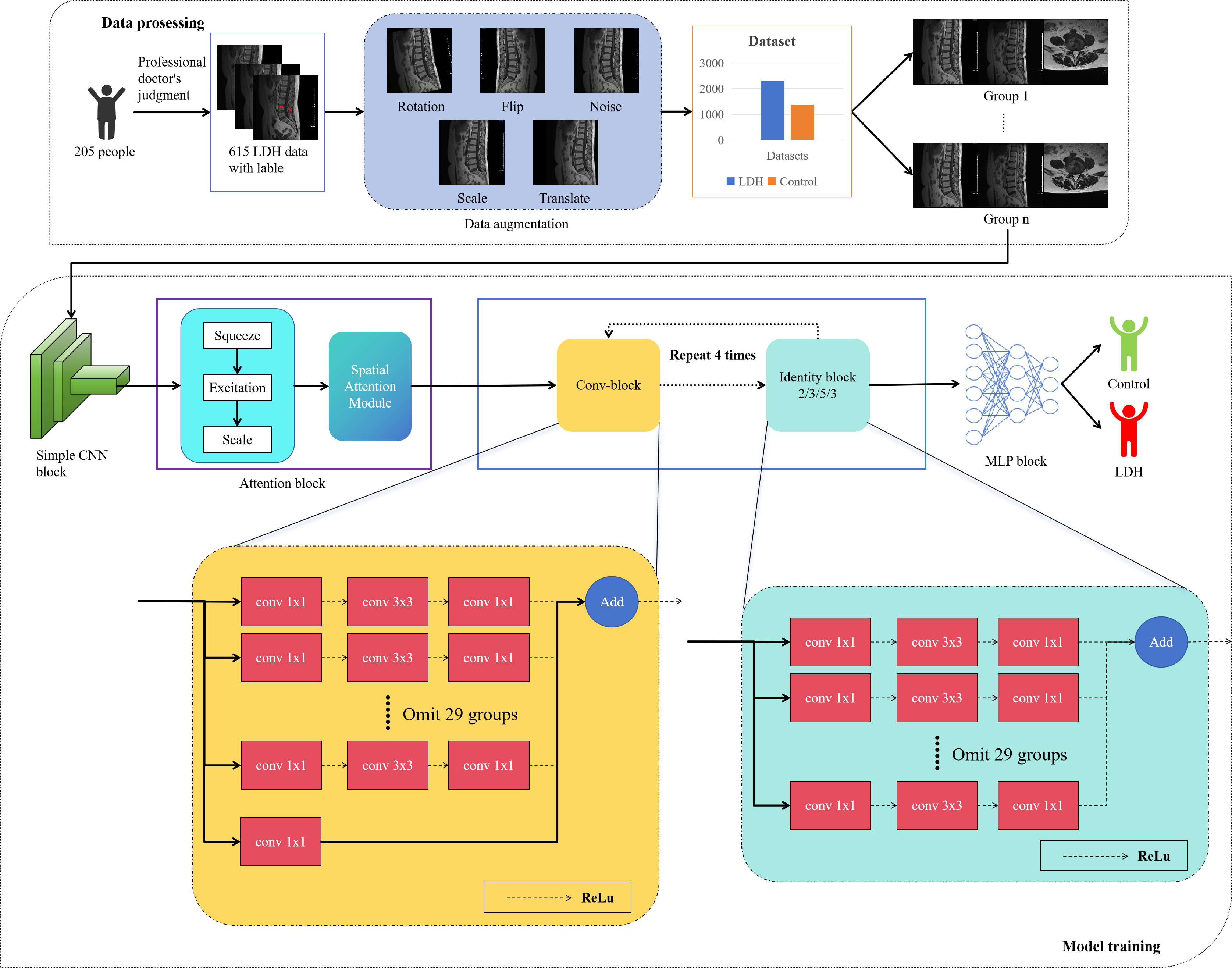}
  \caption{Overview of the structure. The architecture is structured into two primary components: data preprocessing and model training. The preprocessing stage applies data augmentation to MRI images, including rotation, inversion, panning, scaling, and noise injection, yielding 3,070 augmented training samples. Afterwards, the images of the same patient were divided into groups by setting an additional marker. In model training, data is processed through a CNN layer, followed by feature extraction via a dual-attention module integrating channel and spatial attention. Final LDH predictions are generated through a hierarchical network of Conv-block, Identity block, and a Multi-Layer Perceptron (MLP). The Conv-block and Identity block are repeated 4 times, and the Identity block appears 2 times, 3 times, 5 times, and 3 times each time.}
  \label{fig_structure}
\end{figure*}

Researchers are increasingly focusing on developing automated diagnostic systems using deep learning and medical imaging data such as MRIs and CT scans to improve clinical efficiency and accuracy \citep{su2022automatic}. The landscape of deep learning architectures is vast and multifaceted. Notable examples include VGG \citep{simonyan2014very} networks, renowned for their straightforward design employing successive 3×3 convolutional layers, and residual networks (ResNets) \citep{he2016deep}, which revolutionized depth scalability through their innovative residual blocks. These blocks incorporate skip connections that bridge layers, enabling the construction of networks spanning thousands of layers while mitigating gradient vanishing issues. However, a critical limitation of VGG architectures lies in their susceptibility to vanishing gradients as network depth increases, which compromises training stability and convergence efficiency. Conversely, while ResNet architectures achieve remarkable depth through residual learning, their parameter complexity imposes significant computational overhead. This challenge hinders deployment in resource-constrained clinical settings, such as hospital environments relying on edge devices with limited memory and processing capabilities. Riveros et al. \citep{riveros2019comparison} integrated self-organizing mapping (SOM) with K-means clustering to develop a robust diagnostic model. Similarly, Abdullah et al. \citep{abdullah2018prediction} employed Random Forest (RF) and K-Nearest Neighbors (KNN) algorithms to classify LDH, demonstrating the utility of traditional ML approaches in this context. Advances in deep learning have significantly impacted the detection and classification of LDH. Der et al. \citep{nisar2024lumbar} introduced the Novel Lumbar IVD Classification (NLIC) model, built upon the YOLO framework, which underscores the effectiveness of DL in this domain. Furthermore, Sari et al. \citep{sari2024enhanced} combined grey wolf optimization with residual networks to refine feature selection, yielding a more robust classification model. These studies collectively illustrate the transformative potential of ML, particularly DL, in advancing LDH diagnostics.

Data augmentation is a data-centric strategy employed to mitigate the challenges of limited data availability. It comprises a range of methodologies designed to artificially expand the size and diversity of training datasets, thereby enhancing deep learning models' robustness and generalization capabilities. By introducing controlled variations in existing data samples, DA techniques facilitate the development of more reliable and accurate models, particularly in scenarios where data scarcity is a concern \citep{shorten2019survey}. The methodology for data enhancement must prioritize the preservation of data integrity and security, particularly concerning the maintenance of accurate labels post-transformation. This underscores the notion that the efficacy of various data augmentation techniques strongly depends on the specific domain in which they are applied \citep{cubuk2018autoaugment}. 

\section{Methodology}

\subsection{Overview of the structure}

To enhance the clinical relevance and diagnostic utility of model predictions while concurrently emphasizing the detection of nuanced variations within lumbar intervertebral discs, we introduce a novel framework for classifying LDH (Fig.\ref{fig_structure}). This framework is structured into three primary components. A data augmentation strategy is initially applied to the input dataset to expand the sample size available for subsequent model training. Subsequently, set an additional marker by dividing the data from the same patient into groups. This operation allows the model to better learn the relationship between different MRI images. Then, the attention mechanism is used to extract discriminative features from MRI images, enabling the model to prioritize the high-incidence sites of LDH (L4-L5 and L5-S1 lumbar segments). Finally, a deep neural network architecture with ResNeXt as the backbone is used to classify MRI images better to distinguish patients with LDH from healthy individuals.

\begin{figure*}[htbp]
  \centering
  \includegraphics[width=0.9\textwidth]{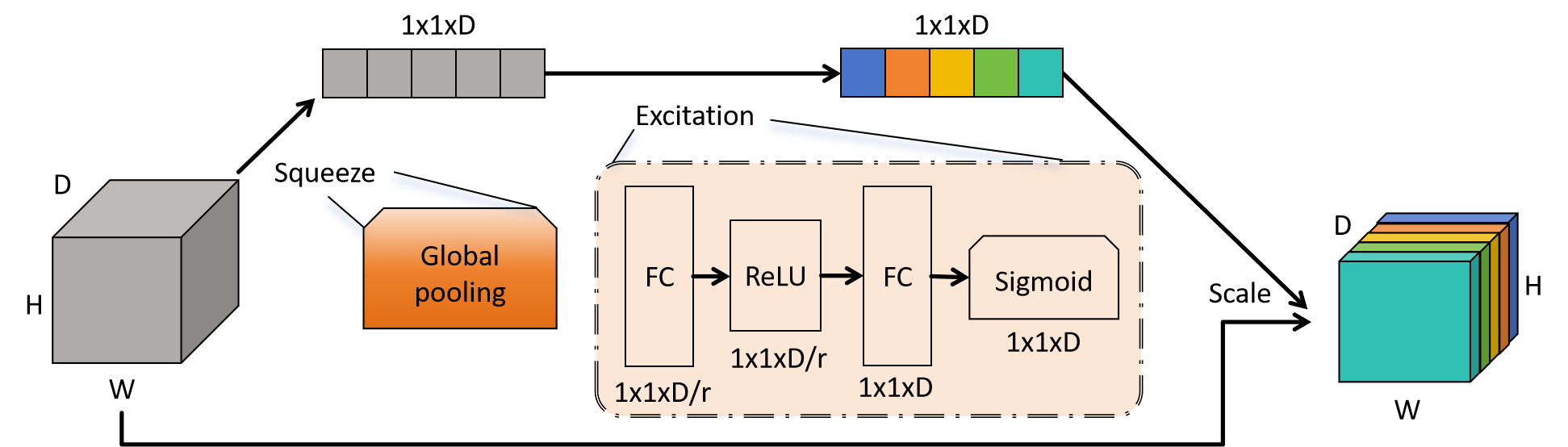}
  \caption{Channel attention block. It obtains the attention features of each channel through three parts: squeeze, excitation, and scale.}
  \label{fig_SEBlock}
\end{figure*}

\subsection{Data augmentation}

We implemented a comprehensive data augmentation strategy to enhance the diversity and robustness of the training dataset for LDH detection. There is variability in MRI acquisition parameters and subtlety in LDH features. The augmentation techniques were carefully selected to simulate real-world imaging variations while preserving critical anatomical details. The augmentation pipeline included:

\textbf{Rotation.} Rotate the image evenly within [-10, 10] degrees. This range avoids anatomical distortion, preserves the spatial relationship between the disc and the nerve root, and does not affect the diagnosis.

\textbf{Filp.} For the MRI images, we focused on using horizontal flips. This augmentation ensures models remain invariant to left-right imaging conventions to enhance their adaptability to the multisite dataset.

\textbf{Translate.} Mimics subtle patient movements during scanning. Translate the image's content horizontally by [-0.1, 0.1] times the side length of the image. This simulates scan positioning deviations and preserves key anatomical structures.

\textbf{Scale.} Zoom in by a factor of [0.8, 1] times the side length of the original image. This simulates different scanning resolutions. It can preserve detail while avoiding image blurring or distortion of critical structures.

\textbf{Intensity-based modifications.} Gaussian noise injection (\(\sigma\) = 0.01-0.05) to simulate low signal-to-noise scans. Within this range, noise does not mask lumbar disc signals, leading to abnormalities.

These augmentations create a more diverse training cohort, reducing overfitting and enhancing model generalization. The constrained augmentation parameters prevent anatomical distortion, ensuring pathological features remain diagnostically meaningful. This balanced approach strengthens model performance and aligns with clinical validation requirements.

\subsection{Model design}

We propose a novel hybrid architecture integrating the Convolutional Block Attention Module (CBAM) \citep{woo2018cbam} into the ResNeXt backbone \citep{xie2017aggregated}. The design aims to enhance feature representation by combining channel-wise and spatial attention mechanisms with the split-transform-merge strategy of grouped convolutions. 

CBAM modules are injected into the residual blocks of ResNeXt to recalibrate feature maps adaptively. We utilize the Squeeze-and-Excitation Block (SEBlock) \citep{hu2018squeeze} to replace the Channel attention block in CBAM. SEBlock is divided into squeeze, excitation, and scale (Fig.\ref{fig_SEBlock}). We perform global average pooling on the input feature map \(x \in R^{c\times H\times W}\) to generate the channel statistics vector \(z \in R^{c\times1 \times1}\) as following:

\begin{align}
    z_{c} = \frac{1}{H\times W}\sum_{i=1}^{H}\sum_{j=1}^{W}x_{c}(i,j)   
\end{align}
Where \(H\) is the height of the input feature map, \(W\) is the width of the input feature map, and \(C\) is the channel.

During the execution of the excitation operation, channel weights are derived by propagating features through two successive fully connected layers. Specifically, this architecture incorporates a ReLU \citep{nair2010rectified} activation layer to introduce non-linearity, followed by a sigmoid layer to normalize the output into the range [0, 1] as follows:

\begin{align}
    s = sigmoid(w_{2} \cdot ReLu(w_{1}z))
    \label{SEblock}
\end{align}
Where \(s\) is the channel attention weight, \(w\) is the fully connected layer's weight matrix for learning the channel relationship.

Finally, the original feature map is dynamically adjusted by channel-by-channel multiplication as \(F\). Important channels are weighted close to 1 to preserve or enhance features; unimportant channels are weighted close to 0 to suppress features as follows:

\begin{align}
    F_{c} = s_{c} \cdot X_{c}
\end{align}
Where \(X_{c}\) is the raw input feature map for channel c.

After implementing the channel attention module, we let the resulting feature map into the spatial attention Module.  spatial attention Module computes spatial attention weights \(M \in R^{1\times H\times W}\) via:

\begin{align}
    M = sigmoid(conv(AvgPool(F)) + conv(MaxPool(F)))
\end{align}
Where \(conv\) represents a convolutional layer.

\begin{figure*}[htbp]
  \centering
  \includegraphics[width=0.9\textwidth]{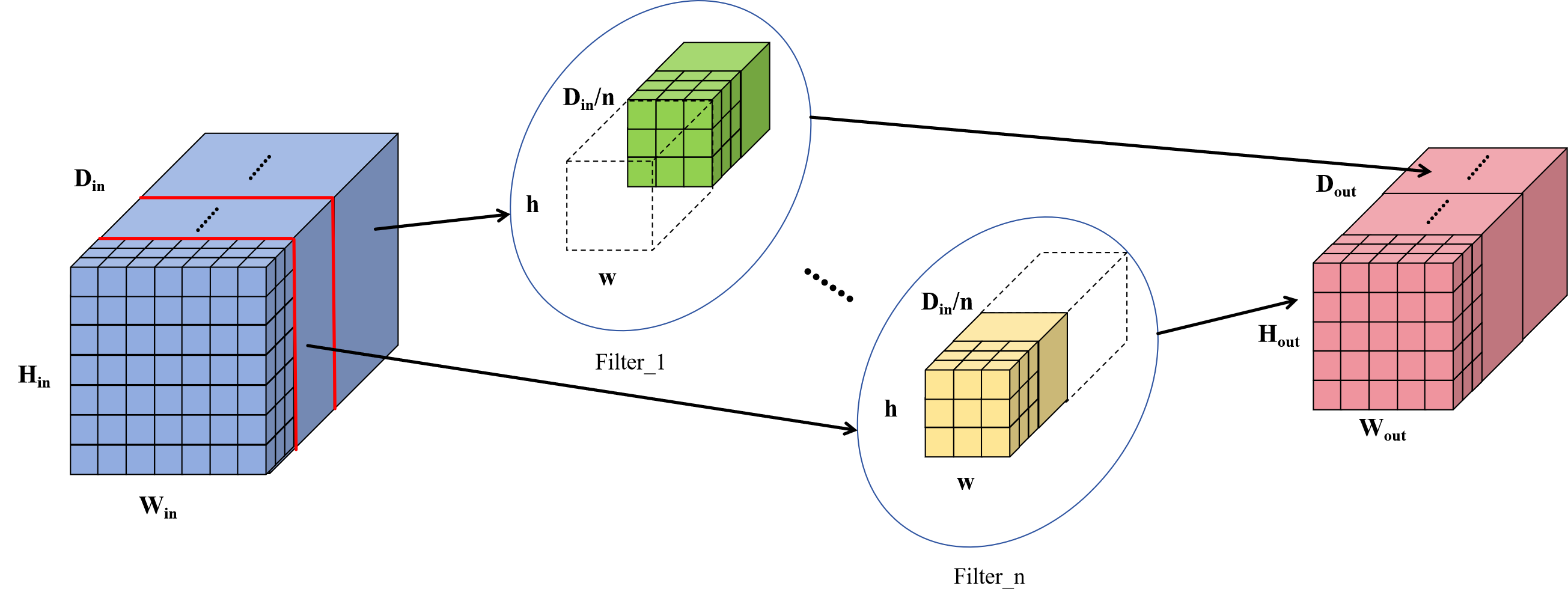}
  \caption{Grouped convolutions used in Conv-block and Identity block. Input features are partitioned into distinct n groups for parallel convolution operations. Finally, the output features are combined.}
  \label{fig_GC}
\end{figure*}

We use a convolutional layer with an output channel of 64, a convolutional kernel of 7, and a step size of 2 as the first layer of the network. We added two attention modules after the first layer of convolution and after the last layer of convolution. The network is then trained using a multilayer module consisting of grouped convolutions (Fig.\ref{fig_GC}) and Residual Networks. Final classification of MRI images by an MLP network. The formula for updating the weights of the MLP is as follows:

\begin{align}
    w_{ji} \gets w_{ji} + \triangle w_{ji} = w_{ji} + \eta \delta x_{ji}
\end{align}
Where \(w\) is the learnable weight, \(x\) is input, \(\eta\)  is learning rate, and \(delta\) is local gradient.

\subsection{Model training}
We train all involved networks on two 4090 servers. We split the dataset into an 80\% training set and a 20\% validation set while using cross-validation for training. The AdamW optimizer was systematically employed throughout the training process due to its demonstrated efficacy in addressing large-scale, non-convex optimization challenges inherent in deep neural networks. This optimizer integrates a decoupled weight decay mechanism that operates as an implicit L2-regularizer, effectively preventing overfitting while maintaining stable convergence in high-dimensional parameter spaces. The updated formula is as follows:

\begin{align}
    \theta_{t+1} = \theta_{t} - \eta (\frac{\hat m_{t}}{\sqrt{\hat v_{t}} + \epsilon}) - \eta \lambda \theta_{t}
\end{align}
\begin{align}
    \hat m_{t} = \frac{m_{t}}{1 - \beta_{1}}
\end{align}
\begin{align}
    \hat v_{t} = \frac{v_{t}}{1 - \beta_{2}}
\end{align}
where \(\theta_{t}\) means model parameters at time step t, \(\hat m\) is the bias-corrected first-order moments of the gradient, \(\hat v\) is the bias-corrected second-order moment of the gradient, and \(\beta\) means exponential decay rate. We set \(\eta\) = 0.001.

\section{Result}

\subsection{Dataset}

\begin{figure}[htbp]
  \centering
  \includegraphics[width=0.4\textwidth]{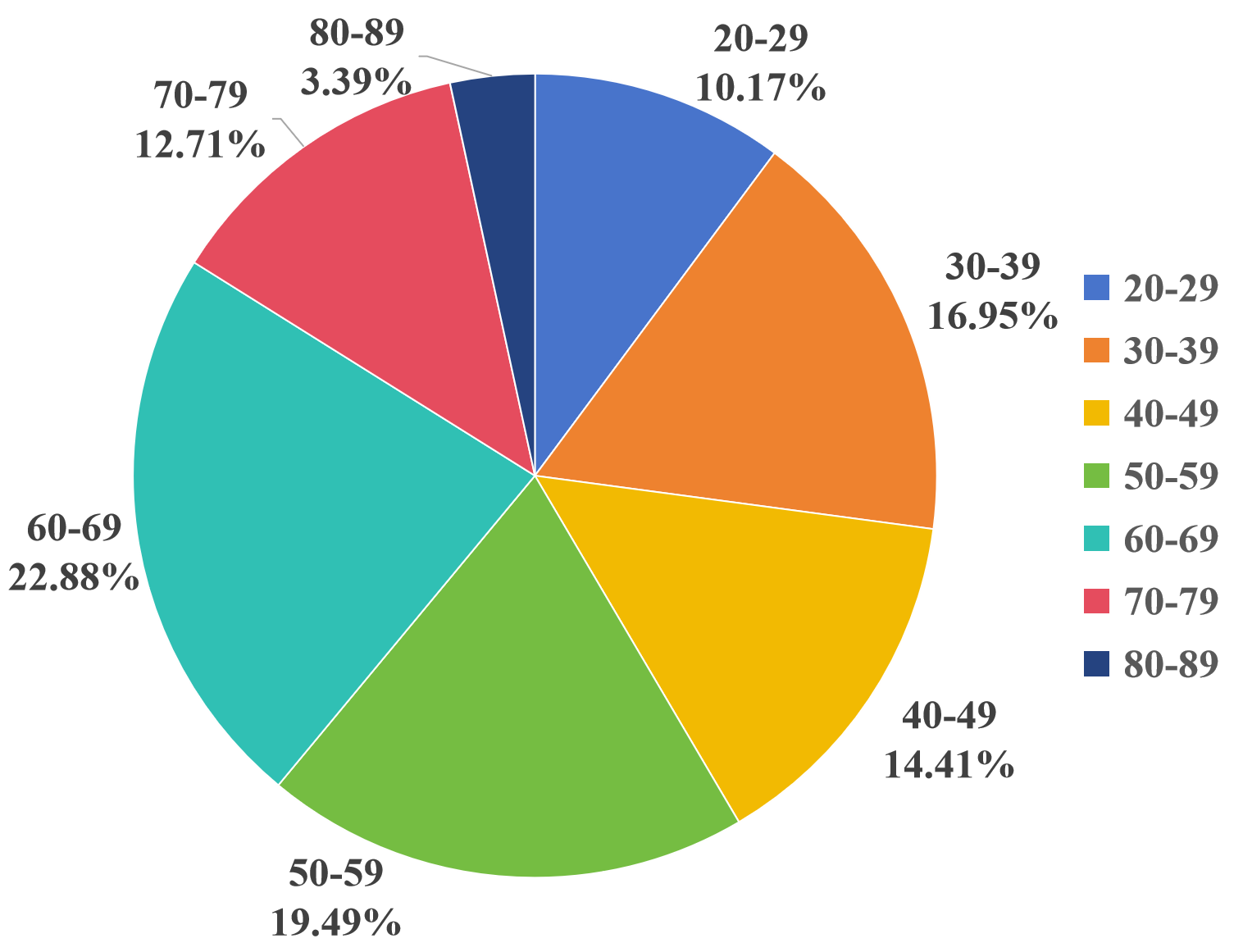}
  \caption{Percentage of different age groups in LDH data.}
  \label{fig_dataset}
\end{figure}

This study compiled a comprehensive dataset on LDH, comprising 615 MRI scans procured from collaborating medical institutions. The dataset includes 387 radiologically confirmed LDH cases diagnosed by expert radiologists based on clinical symptomology and imaging evidence, and 228 healthy control subjects. The control group was meticulously matched to the LDH cohort regarding age, gender, and imaging acquisition parameters to ensure robust comparability. All MRI images were acquired using standardized protocols on identical MRI scanner models and subsequently normalized to a uniform resolution and anatomical orientation, thereby minimizing scanner-related variability. This uniformity facilitates more efficient model training, as the algorithms can focus on learning meaningful radiological patterns rather than adapting to scanner-specific artifacts. Notably, the LDH cohort exhibits a wide age distribution, spanning from 20 to 90 years, which ensures comprehensive coverage across the adult lifespan and captures the diverse clinical presentations of LDH across different age demographics(Fig.\ref{fig_dataset}).

\begin{figure*}[htbp]
  \centering
  \begin{subfigure}[htbp]{0.42\textwidth}
    \includegraphics[width=\textwidth]{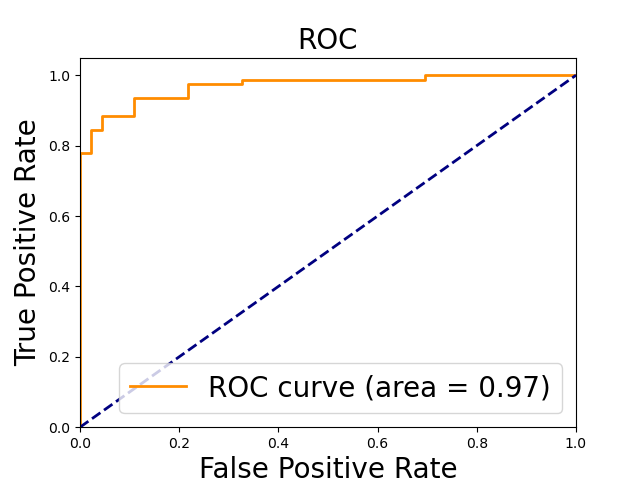}
    \caption{ROC}
    \label{sub_ROC}
  \end{subfigure}
  \hfill
  \begin{subfigure}[htbp]{0.42\textwidth}
    \includegraphics[width=\textwidth]{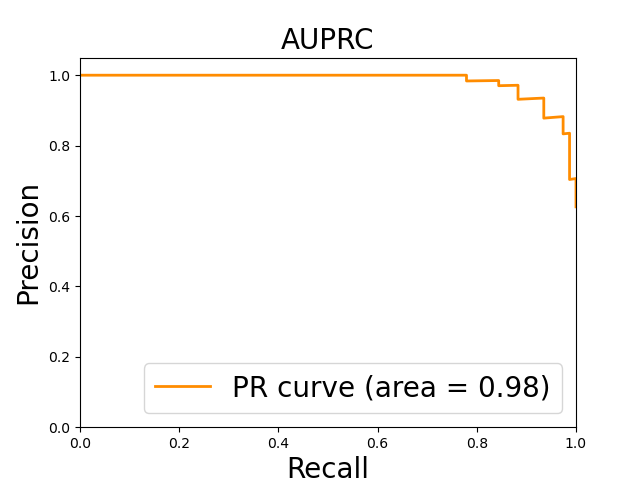}
    \caption{AUPRC}
    \label{sub_AUPRC}
  \end{subfigure}
  \caption{The result of our model. (a) is the result of AUC, (b) is the result of AUPRC.}
  \label{fig_AUC}
\end{figure*}

\begin{table*}[!ht]
    \centering
    \caption{Metrics}
    \label{Tab_metrics}
    \begin{tabular}{c|cccccc}
        Method & ACC & AUC & F1-score & Precision & Recall & AUPRC \\ \hline
        \cite{zhang2023deep} & 0.9286 & 0.915 & 0.9298 & 0.929 & 0.9306 & 0.917 \\
        \cite{sari2024enhanced} & 0.9241 & 0.94 & 0.8984 & 0.8764 & 0.9215 & 0.911 \\
        \cite{kocaman2023investigation} & 0.9225 & 0.939 & 0.847 & 0.8461 & 0.8156 & 0.925 \\
        \cite{tsai2021lumbar} & 0.814 & 0.845 & 0.8951 & 0.8724 & 0.9158 & 0.884 \\
        \cite{wang2024yolov10} & 0.9054 & 0.9351 & 0.9252 & 0.9173 & 0.9333 & 0.956 \\
        Ours & 0.9486 & 0.969 & 0.938 & 0.935 & 0.9412 & 0.981 \\ \hline
    \end{tabular}
\end{table*}

\subsection{Assessment metrics}
We employ a multi-faceted assessment strategy to evaluate our model's performance thoroughly. The Area Under the Receiver Operating Characteristic curve (AUC) quantifies the classifier's overall discriminative ability across all possible thresholds, and the Area Under the Precision-Recall curve (AUPRC) emphasizes accurate positive class identification in imbalanced scenarios. Additionally, visual inspection of the ROC curve provides qualitative insights into the trade-off between true positive and false positive rates. Complementing these, we evaluate performance using threshold-dependent metrics such as Accuracy (ACC), precision, recall, and the F1-score. These metrics were calculated using as following:

\begin{align}
    Recall = TPR = \frac{TP}{TP + FN}
\end{align}
\begin{align}
    FPR = \frac{FP}{TN + FP}
\end{align}
\begin{align}
    ACC = \frac{TP + TN}{TP + TN + FP + FN}
\end{align}
\begin{align}
    Precision = \frac{TP}{TP + FP}
\end{align}
\begin{align}
    F1-score = 2 \cdot \frac{Precision \cdot Recall}{Precision + Recall}
\end{align}
where TP is the number of true positives, TN is the number of true negatives, FP is the number of false positives, and FN is the number of false negatives.

\begin{align}
    AUC = \frac{\sum\textbf{I} (s_T > s_N)  }{n_{T} \cdot n_{N}}
\end{align}
\begin{align}
    AUPRC = \int_{0}^{1} Precision(Recall^{-1}(x))dx 
\end{align}
where \textbf{I} is indicator function, \(s_T\) is predictive scores for positive samples, \(s_N\) is predictive scores for negative samples, \(n_T\) means positive sample size, and \(n_N\) means negative sample size.

\subsection{Detailed results}
The analysis prioritizes the AUC, a pivotal metric for evaluating classification performance. As demonstrated in (Fig.\ref{fig_AUC}), our methodology achieves a superior AUC-ROC of 0.969 (Fig.\ref{sub_ROC}), surpassing all comparative approaches by at least 0.03. And our AUPRC is 0.981 (Fig.\ref{sub_AUPRC}), at least 0.2 higher than all similar methods. This quantitative advantage underscores the model's enhanced discriminative capacity, particularly in distinguishing subtle pathological variations within LDH subtypes.

To make a more comprehensive assessment of the model, we next evaluate the model in terms of other metrics, including ACC, F1 score, precision, and recall. Specific values can be found in (Table \ref {Tab_metrics}). Our approach can get better metrics.

The structural revision implemented in the channel attention module of the CABM within Eq.\ref{SEblock}, yielding a compressed transformation matrix \(w\). Specifically, this modification decomposes the original weight matrix into two lower-dimensional components: \(w1 \in R^{\frac{c}{r} \times c}\) and \(w2 \in R^{c \times \frac{c}{r}}\), where \(r\) is the downscaling ratio. The loss of the maximum pooling branch in the revised architecture reduces the sensitivity to high-frequency texture features compared to the original CBAM design. However, the number of parameters is decreased significantly by 93.75\% when r = 16. This optimization achieves an effective trade-off between feature preservation and parameter savings by maintaining the discriminative power of the model as much as possible while significantly improving the computational efficiency and paving the way for large-scale clinical validation and eventual operationalization.

\section{Conclusion}

This study demonstrates the feasibility and superiority of our model for automated classification of LDH using multi-plane MRI data. Compared to conventional radiological evaluation, the proposed framework achieves high diagnostic accuracy (ACC=0.9486, AUC=0.969) and strong generalizability across anatomical locations. Combining the data augmentation techniques used for magnetic resonance imaging features makes it possible to streamline the workflow and reduce costs.

Our study also has some limitations. Firstly, LDH is characterized by significant heterogeneity, including different types of herniation, comorbidities, and clinical presentations. However, our single data source did not fully reflect this complex disease. Then, the developed framework is specifically tailored to conventional T1-weighted and T2-weighted MRI sequences. It precludes analysis of emerging MRI modalities such as diffusion-weighted imaging or quantitative susceptibility mapping. Finally, the current architecture does not inherently support multimodal data fusion, limiting its capacity to integrate complementary imaging types or clinical covariates that could enhance diagnostic granularity.

These limitations underscore opportunities for future research to expand the model's applicability across heterogeneous imaging ecosystems while maintaining its core diagnostic efficacy. Such enhancements could include adaptive mechanisms for incorporating novel MRI protocols and hybrid architectures capable of cross-modal feature synthesis. And future studies should prioritize the integration of multi-center, diverse datasets that reflect the real-world variability of LDH to verify the model's generalizability.

\section*{Funnding}

This work is supported by National Natural Science Foundation of China under Grant 62473191, Shenzhen Key Laboratory of Robotics Perception and Intelligence (ZDSYS20200810171800001), Shenzhen Science and Technology Program under Grant 20231115141459001, RCBS20221008093305007, Guangdong Basic and Applied Basic Research Foundation under Grant 2025A1515012998, Young Elite Scientists Sponsorship Program by CAST under Grant 2023QNRC001, and the High level of special funds (G03034K003) from Southern University of Science and Technology, Shenzhen, China.

\bibliographystyle{elsarticle-harv} 
\bibliography{reference}

\end{document}